\documentclass[12pt]{article}

\usepackage{adjustbox}
\usepackage{amsmath}
\usepackage{amsfonts}
\usepackage{amsthm}
\usepackage{color}
\usepackage{enumitem}
\usepackage{setspace}
\usepackage{mathrsfs}
\usepackage{amssymb}
\usepackage{bbding}
\usepackage{changepage}
\usepackage{wasysym}
\usepackage{sgame}
\usepackage{natbib}
\usepackage{har2nat}
\usepackage{graphics}
\usepackage{graphicx}
\usepackage{appendix}
\newcommand*{\mybox}[1]{\framebox{#1}}
\usepackage{booktabs,tabularx}

\theoremstyle{definition}

\theoremstyle{plain}

\newtheorem{pred}{Prediction}
\newtheorem{result}{Result}

\theoremstyle{definition}

\theoremstyle{definition}

\theoremstyle{definition}
\newtheorem{theorem}{Theorem}
\theoremstyle{plain}
\newtheorem*{theorem*}{Theorem}
\theoremstyle{plain}

\theoremstyle{plain}
\newtheorem*{corollary*}{Corollary}
\theoremstyle{definition}

\theoremstyle{remark}

\theoremstyle{remark}

\RequirePackage{tabularx}
\RequirePackage{threeparttable}

\newcommand{\abs}[1]{|#1|}

\newcommand{\comment}[1]{}

\newcommand{\set}[1]{\left\{#1\right\}}

\newcommand{\regretCoeff}{\kappa}

\newcommand{\Real}{\mathbb{R}}

\newcommand{\Ui}{u_{i}}
\newcommand{\UiR}{\Ui^{T}}

\newcommand{\Abold}{\mathbf{A}}
\newcommand{\abold}{\mathbf{a}}
\newcommand{\Anoti}{\mathbf{A}_{-i}}
\newcommand{\anoti}{\mathbf{a}_{-i}}
\newcommand{\ai}{a_{i}}

\newcommand{\aihat}{\hat{a}_{i}}

\newcommand{\belief}[1]{\phi_{#1}}

\usepackage[pdftex, citecolor=blue, colorlinks]{hyperref}
\usepackage[margin=3cm]{geometry}
\usepackage{eurosym}
\usepackage{caption}
\usepackage{subcaption}

\begin{document}

%{\title{Ignorance is bliss:\ A game of regret\thanks{Thanks to Zvonimir Ba\^si\'c, Roland B\'{e}nabou, Phil Brookins, Martin Dufwenberg, Christoph Engel, Chris Gee, Andreas Gr\"{u}newald, Senran Lin, Evan Piermont, Antonio Rosato, Juan Pablo Rud, Joel Sobel, Marcel Zeelenberg, and audiences at University of Tilburg (TIBER 2019), UCSD (Southwest Economic Theory conference 2020), UCSB, King's College London and University of Bath. Funding from the Max Planck Society, Middlesex University and Royal Holloway University is gratefully acknowledged.}}}

{\title{Ignorance is bliss:\ A game of regret\thanks{Thanks to Zvonimir Ba\^si\'c, Roland B\'{e}nabou, Phil Brookins, Martin Dufwenberg, Christoph Engel, Chris Gee, Andreas Gr\"{u}newald, Senran Lin, Anne Neary, Evan Piermont, Antonio Rosato,  Frances Ruane, Juan Pablo Rud, Joel Sobel, Marcel Zeelenberg, and audiences at University of Tilburg (TIBER 2019), UCSD (Southwest Economic Theory conference 2020), UCSB, King's College London, and University of Bath. Funding from the Max Planck Society, Middlesex University and Royal Holloway University is gratefully acknowledged.}}}

\author{Claudia Cerrone\protect\footnote{\emph{Email}: \href{mailto:claudia.cerrone@city.ac.uk}{claudia.cerrone@city.ac.uk} \emph{Address}: Department of Economics, City, University London, Northampton Square, London, EC1V 0HB.} \hspace{.4in} Francesco Feri\protect\footnote{\emph{Email}: \href{mailto:francesco.feri@rhul.ac.uk}{francesco.feri@rhul.ac.uk}; \emph{Address}: Department of Economics, Royal Holloway, University of London, Egham, Surrey, TW20 0EX.}  \hspace{.4in} Philip R. Neary\protect\footnote{\emph{Email}: \href{mailto:philip.neary@rhul.ac.uk}{philip.neary@rhul.ac.uk}; \emph{Address}: Department of Economics, Royal Holloway, University of London, Egham, Surrey, TW20 0EX.}}

\date{\today}

\maketitle

%%%% ----------------------------------------------------------------------

\vspace{.3in}

%%% ----------------------------------------------------------------------

\begin{abstract}
%\noindent
\noindent
An individual can only experience regret if she learns about an unchosen alternative.
In many situations, learning about an unchosen alternative is possible only if someone else chose it. 
%This paper begins with the observation that more information after-the-fact is never desirable for regret averse decision makers.
We develop a model where the ex-post information available to each regret averse individual depends both on their own choice and on the choices of others, as others can reveal ex-post information about what might have been.
%regret averse individuals learn about unchosen options with some probability that depends on the decisions of others, as others can reveal ex-post information about foregone alternatives.
%We suppose that there are multiple regret averse individuals and that the ex-post information available to each depends both on their own choice and on the choices of others.
This implies that what appears to be a series of isolated single-person decision problems is in fact a rich multi-player behavioural game, the {\it regret game}, where the psychological payoffs that depend on ex-post information are interconnected. For an open set of parameters, the regret game is a coordination game with multiple equilibria, despite the fact that all individuals possess a uniquely optimal choice in isolation.
%Regret averse players will attempt to coordinate on ex-post information environments.
We experimentally test this prediction and find support for it.

\bigskip

\end{abstract}

%%% ----------------------------------------------------------------------

\setstretch{1.3}
\newpage

\begin{quote}
For of all sad words of tongue or pen, The saddest are these: `It might have been!'
\end{quote}

John Greenleaf Whittier

\section{Introduction}\label{INTRODUCTION}

Anne goes to her local Italian restaurant for dinner.
The restaurant serves only two dishes, spaghetti and risotto.
The spaghetti is always the same and always pretty good.
Ordering it is a safe choice.
The risotto on the other hand is risky.
Sometimes Anne thought it delicious but on occasion it has been tasteless.
Anne suffers from regret:\ if she chooses the risotto and it turns out to be tasteless then she will regret her choice; likewise if she chooses the spaghetti and the risotto turns out to be delicious.
While Anne suffers from regret, she is sufficiently self aware of this trait so that she always acts in a regret-averse manner.
That is, Anne always accounts for the possibility of unpleasant regretful experiences before taking any decision.

%If Ann is regret averse then the possibilities of these unpleasant regretful experiences are important factors to account for in her decision making process.

Information feedback plays an integral role in the above story.
Any regret experienced by Anne involves comparing the utility from the choice made -- in the event it turns out to be suboptimal -- with the utility of the foregone alternative that would have been best with hindsight.
But implicit in this construction is that such a comparison can always be made.
%For a regret averse decision maker choosing between financial assets listed on an exchange, ex-post comparisons are always possible and clearly relevant \cite{KahnemanRiepe:1998:JPM}.
%But there are many situations in life where information about counterfactuals is not readily available ex-post.
While regret averse decision makers can sometimes make ex-post comparisons, for example when choosing between financial assets listed on an exchange \cite{KahnemanRiepe:1998:JPM}, there are many situations in life where information about counterfactuals is not readily available after-the-fact.
And if information about a foregone alternative is not near at hand, then, at least in a model of regret aversion, that foregone alternative cannot be regretted.\footnote{This highlights one way in which models of regret aversion are at odds with everyday usage of the term ``regret'':\ it is not uncommon to hear a person express regret about not taking a particular course of action when in fact they can never know how things would have turned out had they done so.}

%If the ex-post information available does depend on choice,
It is only a small leap to see that allowing the ex-post information available to a regret averse decision maker to vary with choice can impact optimal decision-making.
To illustrate this concretely, let us expand on the above.
% with a more realistic in-built informational asymmetry.
Suppose that Anne is the only patron in the restaurant and that she is not in the habit of ordering more than one main course.
Clearly if Anne orders the spaghetti then she never learns the risotto's quality.
However, since the spaghetti's quality is guaranteed, it can always be benchmarked against whether ordered or not. 
%for example on how it tasted on Anne's last visit to the restaurant -
Thus, there is an implicit in-built information asymmetry since what is known ex-post after ordering the spaghetti differs from that after ordering the risotto.
%Ann is aware of this informational asymmetry.
Can this matter for optimal choice?
The answer is:\ yes, of course it can matter, because differing levels of ex-post information are an important factor to behavioural individuals like Anne.
%matter for behavioural individuals like Ann for whom the available ex-post information is a factor in their decision-making.
In this set up, ordering the spaghetti brings the added advantage that it completely insures Anne against regret, whereas the same is not true of ordering the risotto.
Ignorance is bliss.

While ignorance is bliss for a regret averse decision maker, the problem, as so often in life, is others.
%Often people learn from others.
%People may learn from others.
To see how, let us consider another modification to the above story in which Anne has a friend, let's call him Barry, who joins her for dinner.
%Barry also suffers from regret and is regret averse in exactly the same way as Anne.
%If Barry orders the risotto, then ordering the spaghetti is no longer the safe haven from regret for Ann as it was when she was dining alone.
When dining alone, ordering the spaghetti was a sure safe haven from regret for Anne, but this is no longer true when dining with Barry.
%\marginpar{\tiny{\PN{I think I wrote this footnote. Any idea what I meant? It reads funny to me.}}}
Why? If Barry orders the risotto, then Anne will learn of the risotto's quality irrespective of whether she orders it herself.\footnote{This requires the (hopefully innocuous) assumption that at some point both dining partners will comment on the quality of their respective dishes.}
If Barry's risotto is tasteless, then Anne is unaffected; but if Barry's risotto emerges delicious from the kitchen, then Anne will experience regret and her enjoyment of the spaghetti will be diminished.
%The informational channel is clear:\ 
A new complication materialises because Barry's choice has the potential to impose a negative externality on Anne.
Not in a direct material sense, but merely by ordering the risotto, Barry impacts (probabilistically) Anne's ex-post information, which in turn impacts the total psychological payoff that she associates with ordering the spaghetti.
%Not in a direct material sense, but simply by ordering the risotto, Barry impacts (probabilistically) Ann's ex-post information, which in turn impacts her total psychological payoff (i.e., the payoff from the choice made and the potential psychological loss due to regret) that she associates with ordering the spaghetti.
%Most interestingly in our opinion, if
If Anne is aware of all this, and let us assume that she is, then the fear of missing out (FoMo) on a potentially superior outcome can be so great that when dining with Barry she may order a dish that she would never opt for when dining alone.
%If Ann is aware of this, the fear of missing out (FoMo) can be so great as to induce her to make a choice that abates this fear at the cost of choosing a dish that she would never order if dining alone.
%That is, if Barry ordered first and ordered the risotto, it may be optimal for Ann to do likewise even if she would prefer to order the spaghetti when dining alone.

%On the surface, 

At first glance, a group of diners ordering food in a restaurant appears to be a series of {\it independent} one-person decision problems.\footnote{Our restaurant ordering example is based on one from \citet{ArielyLevav:2000:JCR}. In that paper, the focus is on the differences in patrons' orders in two similar but contrasting settings:\ one where patrons order simultaneously and another where the same patrons order sequentially. \citet{ArielyLevav:2000:JCR} noted disparities in order choice depending on the protocol (in particular the propensity to ``coordinate'').}
But as the Anne and Barry story above shows, upon closer inspection it is not always that.
If the diners are regret averse, then the situation is better-described as an {\it interdependent} behavioural game since the decisions of each can reveal information to others about what might have been.
%\marginpar{\tiny{\PN{this sentence repetitive?}}}
If we assume that Barry is regret averse in exactly the same way as Anne, then ordering the spaghetti is uniquely optimal for either when dining alone.
However, when dining together this may no longer be the case.
Viewed correctly as a game, there are two (pure strategy) equilibria:\ both order the spaghetti and both order the risotto. 
The reason the latter outcome is an equilibrium is that once one orders the risotto, the other cannot escape learning of its quality and so it becomes optimal (i.e, a best response) to do likewise.
Regret aversion is in essence operating as a coordination device, inducing Anne and Barry to coordinate on the same ex-post informational environment.
%Because now the FoMo of a potentially preferred outcome is unavoidable.

%While the way in which varying ex-post information available to a behavioural decision maker might impact optimal choice,
What we believe is the most important point made in this paper is captured by the story above:\ the ex-post information available to an individual may depend not only on her own choice but also (probabilistically) on the choices of others. 
The general topic of how individual psychological motives can lead to socially interdependent decisions is quite underexplored.\footnote{Certainly as compared with that of single-agent informational preferences and with that of standard spillovers in information acquisition due to private signals, e.g., models of herding \cite{BikhchandaniHirshleifer:1992:JPE,Banerjee:1992:QJE} and social learning \cite{DeGroot:1974:JASA}.}
To illustrate formally how this phenomenon can have impact, we define a simple environment - the {\it regret game} - wherein a large group of like-minded regret averse individuals choose from a common binary choice set.
Just as in the story above, one choice is safe and the other risky.
%\marginpar{\tiny{\PN{PN: This seems a bit repetitive to me.}}}
Each individual's choice probabilistically impacts the ex-post information of everyone else.
%We call the environment the {\it regret game}.
% so the the environment is not a series of independent decision problems to be analysed in isolation but rather is in a rich multi-player behavioural game that we term the {\it regret game}.
For an open set of parameters the regret game is a game of coordination that admits multiple equilibria; and this despite the fact that all individuals share a unique optimal choice when faced with the same decision problem in isolation.
Theorem \ref{thm:Nashmany} is a precise statement of this.

%There is also a large and ever-expanding empirical literature on the diffusion of new technologies, and the role that learning plays in such a process (`learning' is another interpretation often afforded to best-response dynamics).  While this issue was noted by sociologists as early as the 1950s (the classic treatment is \cite{Rogers:2003:}), the understanding of econometric issues that arise when analysing social networks, specifically the ``reflection problem'' \citep{Manski:1993:RES}, has led to a recent explosion of work in the area.  To name but a few, \cite{BandieraRasul:2006:EJ}, \cite{ConleyUdry:2010:AER}, and \cite{BanerjeeChandrasekhar:2013:S} explore how the behaviour of one's neighbours affects the decision to adopt a new crop, a new technology, or whether of not to participate in a microfinance programme respectively.  

%\marginpar{\tiny{\PN{PN: I tweaked this. Feel free to change.}}}
An important real life phenomenon that the regret game can help explain is the slow adoption of new and promising technologies (e.g., agricultural technology, green technology, information and communication technology, and so on).
In the context of agriculture, farmers are often slow to update to new crop technologies (the classic study is \citet{RyanGross:1943:RS}).
More recently, in the context of developing countries, \citet{BandieraRasul:2006:EJ}, \citet{ConleyUdry:2010:AER}, and \citet{BanerjeeChandrasekhar:2013:S} documented how the behaviour of one's neighbours affects the decision to adopt a new crop, a new technology, or whether or not to participate in a microfinance programme respectively.
Similar to our restaurant example, farmers face a choice between a safe option (e.g., the existing technology) and a risky option (e.g., the new technology).
Even if the expected productivity of the new technology is higher than that of the existing technology, regret averse farmers may stick to the existing technology to avoid potential regret.
If everybody sticks to the existing technology, nobody will ever learn whether the new technology was successful or not.
In order for it to become the new dominant standard, not only must the new technology turn out successful, but enough others must adopt it so that this fact becomes widely known.

The key feature of our set up is that individuals need not learn the outcome of foregone alternatives. 
This same observation was made in \citet{Humphrey:2004:JEP} who noted that if the outcome of the chosen lottery is revealed and this outcome was possible in two or more states, then the decision maker does not know the state with certainty. 
%This has consequences for a regret averse decision maker since she considers every possible state and ``zeroes-in'' on the best-performing lottery in each, creating a state-dependent reference point for each state that can be matched but never exceeded.
As Humphrey notes, if the decision maker can only be sure that she is in one of many possible states, then she does not know which alternative to benchmark against. 
Humphrey extended the classic decision theoretic models of regret \citep{Bell:1982,LoomesSugden:1982:EJ} to account for this fact, whereby a coarsening of ex-post information occurs for lotteries that pay the same in multiple states.
Our set up shares the same feature since by definition the risk free lottery pays the same in every state.
However, a coarsening of the ex-post informational environment is not guaranteed due to the choices of others.\footnote{In fact, if other individuals randomise, then there will be an additional probability distribution over what ex-post information environment will be faced.}

Some existing multi-agent models share features similar to our regret game.
%\citet{Filiz-OzbayOzbay:2007:AER} incorporate anticipated regret into the preferences of agents partaking in a sealed-bid auction, and demonstrate that in first-price auctions overbidding with respect to the risk neutral Nash equilibrium might be driven by anticipated ``loser regret'' (felt when bidders lose at an affordable price).
\citet{Filiz-OzbayOzbay:2007:AER} demonstrate that overbidding with respect to the risk neutral Nash equilibrium in a first-price sealed-bid auction might be driven by anticipated ``loser regret'' (felt when bidders lose at an affordable price).\footnote{Unlike in our paper, the informational environment does not depend endogenously on the behaviour of the auction participants. Moreover, the ex-post information environment is common to everyone whereas our model allows for the possibility that not all individuals have the same ex-post information.}
%\citet{Benabou:2013:RES} considers the harmful issue of ``groupthink''.
Using preferences for late-resolution of uncertainty from \citet{KrepsPorteus:1978:E}, \citet{Benabou:2013:RES} presents a dynamic model that addresses the harmful issue of ``groupthink''.
Each individual's payoff depends on the effort level of everyone (including his own) and the realisation of a random variable.
%The key feature is the inclusion of \emph{anticipatory utility} - utility that is experienced from thinking about one's future prospects.
%The more positive the forecast, the better for the individual.
The model allows for multiple equilibria, including one in which everyone in the population collectively ignores a negative public signal about the random variable.
%Such collective delusions can persist because individual $j$'s informational decision and effort choice can affect individual $i$'s psychological payoff, leading $i$ to make a different informational decision than he otherwise would.
%In our model by contrast, $j$'s choice can impact the informational environment that $i$ faces, thereby changing $i$'s psychological payoffs and leading $i$ to make a different risk-taking decision than he would make in isolation.
\citet{BattigalliDufwenbergLi} use the {\it psychological games} framework of \citet{GeanakoplosPearce:1989:JET} to incorporate regret preferences into extensive form games.
Payoffs depend not only on behaviour but also on beliefs, and their set up can allow for ``what might have been'' to be payoff relevant.
\citet{CooperRege:2011:GEB} present a model of ``social regret'' wherein the regret from a choice that turned out suboptimal is dampened if others chose the same. 
%The more likely another individual is to choose an alternative, the greater the expected regret from not choosing that alternative. 
That is, the effect of social regret is magnified by the number of others who make a different choice, and less intense when others behave likewise. Experimental results confirm their hypothesis that social regret is a powerful force.\footnote{In contrast, in our model an individual's choice imposes an externality on others by changing the information environment that they may face.}

With the theoretical result pinned down, we test the predictions of the regret game through an experiment.
Our goal is to determine whether regret averse participants attempt to coordinate with their partner as our model predicts.
However, this objective faces an obstacle as our theory holds only for subjects that are regret averse.
As such, we first screen for such individuals via a novel procedure:\ all participants are asked to choose between a sure amount of money and a risky lottery.
Participants who choose the sure amount are asked whether they want to learn the lottery's outcome or not, where avoiding information comes with a small effort cost.
Those who choose not to learn the lottery's outcome and pay the associated cost are classified as {\it regret averse}, whereas those who choose to learn are classified as {\it non regret averse}.
%We note that, to the best of our knowledge, it is not standard for experimenters to inform subjects of the outcomes of foregone alternatives after the fact.

% that, as our decision-theoretic framework shows, this can be a decisive factor for optimal choice.
%We discuss this further in Section~\ref{ssec:linearWarning}.}

%regret averse participants with non regret averse participants.
%Participants were informed about their partner's previous decisions.
%, as our model assumes that the environment is common knowledge. 
%Regret averse participants play a variant of the regret game.
In the second part of the experiment we pair up participants.\footnote{In our initial sessions, we pair participants like-with-like:\ regret averse with regret averse and non regret averse non regret averse. In later sessions, we allow for pairs of different types.}
Again, all participants must decide whether they want to learn the lottery's outcome.
But now there is a key difference:\ a subject will not learn the outcome of the risky lottery only if they choose not to and their partner decides the same.
%Non regret averse participants face the same decision, but their decision environment is different from that of regret averse participants, as it is a dominant strategy for them to learn the lottery's outcome.
%After making their choice, we ask all participants to guess their partner's decision.
According to our theory, this additional feature should matter for regret averse participants but not for those who are not.
This is because it is a best-response for a regret averse individual to choose the action that they believe their partner chose (i.e., to try and coordinate), whereas non regret averse participants should choose to learn the lottery's outcome independent of their beliefs about their partner's decision (as doing so is a dominant strategy for them).

%coordinate with  since it remains a dominant strategy for them to learn the foregone lottery's outcome.
%According to our theory, it is a best response for a regret averse individual to coordinate with their partner. Thus, we expect regret averse participants to choose the action that they believe their partner chose. In contrast, we expect non regret averse participants to choose to learn the lottery's outcome independent of their beliefs about their partner's decision, as it is a dominant strategy for them to learn.

We find support for these predictions.
Regret averse participants' modal action depends on their beliefs about their partner's choice demonstrating a strong propensity to try and coordinate with their partner. 
When regret averse individuals believe that their partner chose not to learn the foregone outcome, nearly all of them (92\%) choose the same.
When regret averse individuals believe that their partner chose to learn the foregone outcome, a large fraction of them (43\%) behave likewise, despite the fact that learning is not what they chose in isolation. 
The modal action of non regret averse participants does not vary with belief.

%whereas non regret averse participants' modal action does not.
%% depend on their beliefs about their partner's choice.
%%When they believe that their partner chose not to learn, nearly all regret averse participants also choose not to learn.
%That is, regret averse participants demonstrate a strong propensity to try and coordinate with their partner.
%\marginpar{\tiny{\PN{Isn't this sentence repetitive?}}}
%In particular, when they believe that their partner chose to learn the foregone outcome, a large fraction of regret averse participants behave likewise, despite the fact that learning is not what they chose in isolation. 
%%his indicates that they switched from their preferred option to the alternative option in the attempt to coordinate with their partner.

We conclude the introduction with an observation.
The term {\it asymmetric information} is typically used to describe any situation where, in advance of some economic interaction, 
%say a potential exchange, 
one party is relatively more informed than their counterpart.
%Identifying such settings is important because of how market outcomes may differ from those wherein everyone has the same information \citep{Akerlof:1970:QJE}.
But it is important to note that these settings are more accurately classified as possessing ``ex-ante asymmetric information'' as there is a clear temporal component:\ initially the parties have differing levels of information, whereas upon conclusion of any transaction the less-informed party is often more informed.
The regret game has, in a sense, the exact opposite feature of ``ex-post asymmetric information''.
At the onset, all regret averse individuals possess the same information; it is only after choices have been made that differing levels of information may exist.
%Our regret game shows that a
A new type of market failure can emerge in settings of this kind:\ all individuals take a decision that is collectively suboptimal so as to avoid being the one with more, not less, ex-post information.\footnote{In environments with ex-ante asymmetric information, it is well known that natural mechanisms may arise that allow the market to ``correct'' for existing market failures, e.g., signaling \cite{Spence:1973:QJE}, screening \cite{RothschildStiglitz:1976:QJE}, etc. It is interesting to consider what sort of mechanisms would correct for the sorts of market failure that arise due to varying ex-post information. We believe this is important for policy makers to be aware of.}

The structure of the paper is as follows.
Section~\ref{sec:game} defines the regret game and classifies the equilibrium set.
Section \ref{sec:experiment} describes the design and results of an experiment aimed at testing the regret game's predictions.
Section \ref{sec:conclusion} concludes.

\section{The Regret Game}\label{sec:game}

Ex-post information feedback is an integral part of the regret story.
If the outcome of a foregone alternative is never learnt, then how can it be regretted?
Answer:\ it can't.
In this section we suppose that the ex-post informational environment faced by a collection of regret averse decision makers depends not only on own choice but also potentially on the choices of others.
This means that what may at first appear to be a series of single-person decision problems is in fact an interdependent behavioural game that we term the {\it regret game.}
In Section~\ref{ssec:theGame} we define the regret game, and in \ref{ssec:equilibria} we classify the equilibrium set.

\subsection{The regret game}\label{ssec:theGame}

There is a state space $\Omega = \set{\omega_{1}, \omega_{2}}$, where $\omega_1$ occurs with probability $\alpha \in (0, 1)$ and $\omega_2$ with probability $1-\alpha$.
We define a symmetric $N$-player (simultaneous-move) behavioural game with common binary-action set.
%For every player $i$, their utility, $\UiR$, is made up of a real-valued choiceless utility function $u$ and a regret term 
% and common utility function $u^{T}$.
%We identify the action set $A$ with the 2-element choice set $\Lcal = \set{\ell, \ell_{S}}$.
Each player $i$ chooses an action $\ai \in A_i = \set{\ell, \ell_{S}}$, where $\ell$ is a risky lottery that pays $\bar{\ell}$ in state $\omega_1$ and $\underline{\ell}$ in state $\omega_2$, and $\ell_{S}$ is a safe risk free lottery.
We assume that $\underline{\ell} < \ell_S < \bar{\ell}$.
Each individual has utility function $\UiR : \Abold \times \set{\omega_{1}, \omega_{2}} \to \Real$, where $\Abold := \prod_{j =1}^{n} A_{j}$.
We write $\abold = (a_{1}, \dots, a_{n})$ to denote a typical element of $\Abold$ and refer to such objects as action profiles.
From player $i$'s perspective, an action profile $\abold $ can be viewed as $(\ai, \anoti)$, so that $(\aihat, \anoti)$ will refer to the profile $(a_{1}, \dots, a_{i-1}, \aihat, a_{i+1}, \dots, a_{n})$, i.e., the action profile $\abold$ with $\aihat$ replacing $\ai$.

The structure of the ex-post information environment is as follows.
The risk-free lottery, $\ell_S$, has the property that its outcome will always be known ex-post whether it was chosen or not.
The same need not be true of the risky lottery, $\ell$.
For each player $i$, we let $\belief{i} : \Anoti \to [0,1]$ be a function that specifies the probability that player $i$ learns the outcome of the risky lottery, $\ell$, conditional on choosing the safe lottery, $\ell_{S}$.
We emphasise this key modelling assumption:\ when player $i$ chooses the safe option, the likelihood that she then learns the outcome of the risky lottery depends on the behaviour of others, $\anoti$.
To ensure that the environment is symmetric, we further assume that function  $\belief{i}$ is the same for all players $i$, so that it will be denoted by $\belief{}$.\footnote{This involves an abuse of notation as the domain of $\belief{}$ differs for each player.}

Given all this, the utility of player $i$ can be defined as in \eqref{eq:UiRstate0} and \eqref{eq:UiRstate1} below.
% one difference:\ the probability that $i$ learns the outcome of the risky lottery when it is not chosen, denoted $\qi$, depends on the behaviour of the other agents.
\begin{equation}\label{eq:UiRstate0}
\UiR\big((\ell_{S}, \anoti), \omega\big) = \left\{ \begin{array}{rl}
 u(\ell_{S}) - \belief{}(\anoti)\regretCoeff\big(u(\overline{\ell}) - u(\ell_{S}) \big), &\mbox{ if } \omega = \omega_{1} \\
 u(\ell_{S}), &\mbox{ if } \omega = \omega_{2}
       \end{array} \right.
       \end{equation}
and
\begin{equation}\label{eq:UiRstate1}
\UiR\big((\ell, \anoti), \omega\big) = \left\{ \begin{array}{rl}
 u(\overline{\ell}) , &\mbox{ if } \omega = \omega_{1} \\
 u(\underline{\ell}) - \regretCoeff\big( u(\ell_{S}) - u(\underline{\ell}) \big), &\mbox{ if } \omega = \omega_{2}
       \end{array} \right.
\end{equation}
where $u$ is a common choiceless utility that comes directly from the chosen lottery and, where present, the second term captures the regret experienced when the chosen lottery turned out not to be optimal.\footnote{The term ``choiceless utility'' is due to \citet{LoomesSugden:1982:EJ}. It is the utility if the DM is assigned a lottery, without actively choosing it.}
We have assumed a linear regret penalty term where all players have the same non-negative coefficient of regret aversion, $\regretCoeff$.
Assuming that the regret penalty is linear is a simplification that can be relaxed. However, doing so comes at great notational cost, and given that our focus is on the role of informational feedback as a result of the behaviour of others plays in decision making and not on the specific form of regret, we deemed this specification worth it as it allows us to explore this feature in a tractable way.
For a similar reason we assume that individuals experience regret, and we rule out experiencing the reciprocal emotion of rejoice.

We close the model by assuming that each player's probability of learning the outcome of $\ell$ conditional on choosing $\ell_S$, $\belief{}(\anoti)$, is strictly increasing in the number of others that choose $\ell$.\footnote{This assumption seems natural to us. The more individuals who choose the risky lottery, the harder it ought be not to learn of its outcome ex-post. However, while it seems implausible to us that the function $\belief{}$ would be decreasing at any point, we allow for a general functional form. One can imagine settings in which $\belief{i}$ is linear, concave, convex, etc. Relaxing the assumption that $\belief{}$ is strictly increasing, one can envisage settings where $\belief{}$ is step function, in that the outcome of $\ell$ cannot be avoided once {\it enough} individuals have chosen it.}
That is, abusing notation somewhat, we let $\abs{\anoti} := \#\set{j \neq i : a_{j} = \ell }$.
Then for any two profiles $\abold'$ and $\abold''$, player $i$ finds $\belief{}({\abold'}) > \belief{}({\abold''})$ if and only if $\abs{\anoti'} > \abs{\anoti''}$.
Finally, we assume that if nobody chooses the risky lottery then its outcome is learned by noone, and that if everyone other than you chooses the risky lottery, then you cannot escape learning of its outcome.
These conditions are captured formally by $\belief{}(0) = 0$ and $\belief{}(N-1) = 1$.

Both the risky lottery, $\ell$, and the risk-free lottery, $\ell_{S}$, bring with them a benefit and a cost.
The benefit is the direct choiceless utility associated with each; the cost is the psychological penalty that may be incurred in the event your choice is not optimal in the realised state.
For the risky lottery, $\ell$, both the expected benefit and the expected cost are fixed.
For the risk-free lottery, $\ell_{S}$, the expected benefit is fixed but the expected cost is not. The likelihood of a cost being experienced depends on the behaviour of others.
{Thus, we note the somewhat curious feature that choosing the risky lottery brings with it a guaranteed (expected) utility, whereas choosing the risk-free lottery yields an expected payoff that is not locked in as it varies depending on the decisions of others.}

We emphasise that, while player $i$'s utility depends on the choices of everyone, and hence the set up is a strategic game, the dependence is not direct. Rather the dependence manifests through the likelihood that player $j$'s choice of lottery will impact the ex-post information available to player $i$.
That is, individual $j$'s risk taking can generate information for individual $i$ that in turn alters the psychological payoffs that $i$ associates with each choice.

Using \eqref{eq:UiRstate0} and \eqref{eq:UiRstate1}, it is straightforward to calculate when the risky lottery is preferable to a regret averse individual.
To make things as clear as possible, we normalise the choiceless utility function $u$ so that $u(\underline{\ell}) = 0$ and $u(\ell_{S}) = 1$. With this, the condition for the risky lottery $\ell$ to be preferred to the risk-free lottery $\ell_{S}$ reduces to the following threshold rule
\begin{equation}\label{eq:EURcondition}
u(\overline{\ell}) \geq \frac{1}{\alpha}\left(\frac{1+\regretCoeff \big(1-\alpha(1-\belief{})\big)}{1+\belief{}\regretCoeff}\right)
\end{equation}

The expression in \eqref{eq:EURcondition} is bookended by two important cases.
When $\phi=1$ the decision maker will learn the outcome of the risky option no matter what.
Since the regret penalty term is linear, it is straightforward to show that there is no distortion to the threshold rule relative to standard preferences (i.e., when utility is given by choiceless utility only, or equivalently when $\regretCoeff = 0$).
In this case we get,
\begin{equation}\label{eq:EURconditionq=1}
u(\overline{\ell}) \geq \frac{1}{\alpha}
\end{equation}

When $\belief{} = 0$ however, regret averse individual $i$ knows that she will definitely \emph{not} learn the outcome of the risky option unless he chooses it herself.
Then \eqref{eq:EURcondition} becomes,
\begin{equation}\label{eq:EURconditionq=0}
u(\overline{\ell}) \geq \frac{1}{\alpha}\big( 1+\regretCoeff(1-\alpha) \big)
\end{equation}

Given that $\alpha \in (0, 1)$, whenever $\regretCoeff>0$ the inequality in \eqref{eq:EURconditionq=0} is a more demanding condition on the risky lottery $\ell$ than that in \eqref{eq:EURconditionq=1}.
This means that when a regret averse decision maker will certainly not learn the realisation of the risky lottery without choosing it, they require it to have a more desirable payoff distribution.\footnote{It can be checked that the RHS of expression \eqref{eq:EURcondition} is strictly decreasing in $\belief{}$ over the interval $[0, 1]$.}
If $u(\overline{\ell})$ is in the interval $\big[\frac{1}{\alpha},\frac{1}{\alpha}\big( 1+\regretCoeff(1-\alpha) \big)\big]$, then the risk free lottery is optimal when the outcome of the risky lottery is learned only if chosen, while the risky lottery is optimal otherwise.

The reason for the discrepancy above is that the risk-free lottery can provide insurance against regret.
When $\phi=0$, the risk-free lottery provides complete insurance against regret, since there is asymmetry in anticipated regret:\ if the decision maker chooses the risky option, he knows he will be able to make an ex-post comparison and feel regret if the risky option is not successful, whereas if he chooses the risk-free lottery, he knows he will not be able to make such a comparison. Because of the insurance against regret offered by the risk-free lottery, the outcome of the risky lottery in the event of a success must be high enough to tempt the decision maker away from the security of the risk-free lottery. Ignorance is bliss. On the other hand, when $\phi=1$, the risk-free lottery offers no insurance against regret. The decision maker knows he will learn the outcome of the risky lottery whether he chooses it or not. The regret considerations are symmetric and cancel each other out, and the individual's condition is the same as if he were regret neutral ($\regretCoeff = 0$).

Intuitively, the benefit that the risky lottery must yield in order to be chosen is decreasing in $\phi$ due to the reduction in anticipated regret.
To put it another way, as $\phi$ decreases (i.e., the likelihood of making an {ex-post} comparison in the case the risky option is not chosen reduces) an individual is increasingly - from an {ex-ante} perspective - insured against regret.  And because of this insurance against potential regret, the risky lottery's outcome must increase to tempt the agent away from the safe option.

We now characterise the set of equilibria to this set up.

\subsection{Equilibria in the regret game}\label{ssec:equilibria}

The above defines the regret game.
There are three classes of the regret game depending upon the parameters.
When $u(\overline{\ell})<\frac{1}{\alpha}$, it is a dominant strategy for each player to choose the safe lottery $\ell_{S}$.
Similarly, when $u(\overline{\ell}) > \frac{1}{\alpha}\left( 1+\regretCoeff(1-\alpha) \right)$, choosing the risky lottery $\ell$ is optimal no matter how others behave.
For intermediate values of $u(\overline{\ell})$ however, the game is one of coordination and has exactly two pure-strategy Nash equilibria:\ all players choose $\ell_{S}$ or all players choose $\ell$.\footnote{There is also a completely mixed strategy equilibrium but we ignore it as it is highly unstable.}
Theorem \ref{thm:Nashmany} below states this formally.

\begin{theorem}\label{thm:Nashmany}
In the regret game:
\begin{enumerate}
\item\label{condition:Nashmany1}
%$\set{(\ell_{S}, \dots, \ell_{S})}$, when $u(\overline{\ell}_{r}) < \frac{1}{p}$,
When $u(\overline{\ell}) < \frac{1}{\alpha}$, uniform adoption of the risk-free lottery, $\ell_{S}$, is the unique (dominant) pure strategy Nash equilibrium.

\item\label{condition:Nashmany2}
%$\set{(\ell_{r}, \dots, \ell_{r})}$, when $u(\overline{\ell}_{r}) > \frac{1}{p}\left( 1+k(1-p) \right)$,
When $u(\overline{\ell}) > \frac{1}{\alpha}\left( 1+\regretCoeff(1-\alpha) \right)$, uniform adoption of the risky lottery, $\ell$, is the unique (dominant) pure strategy Nash equilibrium.

\item\label{condition:Nashmany3}
%$\set{(\ell_{S}, \dots, \ell_{S}),(\ell_{r}, \dots, \ell_{r})}$, when $u(\overline{\ell}_{r}) \in [\frac{1}{p}, \frac{1}{p}\left( 1+k(1-p) \right) ]$.
When $u(\overline{\ell}) \in \big[\frac{1}{\alpha}, \frac{1}{\alpha}\left( 1+\regretCoeff(1-\alpha) \right) \big]$, uniform adoption of the risk-free lottery, $\ell_{S}$, is a pure strategy Nash equilibrium and uniform adoption of the risky lottery, $\ell$, is a pure strategy Nash equilibrium. Moreover, there are no other pure strategy Nash Equilibria.

\end{enumerate}
\end{theorem}

A proof of Theorem \ref{thm:Nashmany} can be found in Appendix~\ref{App:proofs}.
The result can be understood as follows.
While the risk-free lottery, $\ell_{S}$, brings a fixed choiceless utility, for every additional individual who chooses the risky lottery, the expected total utility associated with the risk-free lottery is decreasing.
The reason for this being that the likelihood of learning about the alternative, and hence experiencing regret, is only going up.
Thus, for the parameters of case \ref{condition:Nashmany3}, eventually there will be a threshold, or tipping point, at which ``enough'' others are choosing the risky lottery that it becomes optimal to follow suit.
That is, for the parameters of case \ref{condition:Nashmany3}, the environment is a coordination game.
However, we note that it is a slightly unusual coordination game in that the number of others who choose $\ell$ is decreasing the associated net utility of choosing the other option $\ell_{S}$ without improving the net utility of choosing $\ell$.

Given that the regret game is one of coordination for the parameters of case \ref{condition:Nashmany3} of Theorem~\ref{thm:Nashmany}, we include a brief discussion of welfare and equilibrium selection.

\paragraph{Welfare and equilibrium selection}
From \eqref{eq:UiRstate0} and \eqref{eq:UiRstate1}, one can compute that uniform adoption of the risky lottery, $\ell$, is Pareto optimal if and only if $u(\overline{\ell}) \geq \frac{1}{\alpha}\big( 1+\kappa(1-\alpha) \big)$.

But this is precisely the threshold above which, for a regret averse individual, the adoption of the the risky lottery is a dominant strategy. Thus, for parameters such that the regret game is a coordination game, coordinating on the risk-free lottery is Pareto-optimal (i.e., preferred ex-ante).

As regards which of the two equilibria is most likely to be observed, we can appeal to the large literature addressing the multiplicity issue that occurs in coordination games.
The most commonly considered environment is a large-population binary-action game (e.g., the Stag Hunt) where the Pareto efficient equilibrium is risky and does not coincide with the {safe} risk-dominant equilibrium.\footnote{A well known experimental test of the efficiency versus safety issue is \citet{Van-HuyckBattalio:1990:AER}, who study behaviour in the \emph{minimum effort game}, of which the Stag Hunt is a simple binary-action version (see \citet{Crawford:1991:GEB}).}
Existing equilibrium selection techniques - be they evolutionary like {\it stochastic stability} \citep{FosterYoung:1990:TPB,KandoriMailath:1993:E, Young:1993:E} or higher-order belief-based like {\it global games} \citep{CarlssonDamme:1993:E, MorrisShin:2003:} - favour the equilibrium that is more difficult to destabilise (i.e., the risk-dominant one).
In the regret game, precisely where the threshold for switching occurs will depend on the functional form of $\belief{}$, which will in turn determine the selected outcome.

With further parameterisation of the model, specifically by fixing values for $\alpha, \kappa$, and $u(\overline{\ell})$, it is possible to compute what values of $\belief{}$ will the risky lottery be most desirable.
(Recall that an increase in $\belief{}$ does not impact the utility of the risky lottery but strictly lowers that of the risk-free lottery.)
There are parameterisations such that $\belief{i}(\abs{\abold}) \approx 1$ whenever $\abs{\anoti} \geq 1$, for which uniform adoption of the risky lottery would be the predicted outcome (since adoption of the risk-free lottery is so unstable).
In this case, the regret game has the interesting property that the common choice of the risk-free lottery, that is also the Pareto dominant outcome, need not be risk-dominant as per \citet{HarsanyiSelten:1988:}.

\section{Experiment}\label{sec:experiment}

In this section we describe the experiment that tests whether regret averse individuals coordinate as our model predicts.
Section~\ref{sec:design} describes the experiment, Section~\ref{sec:predictions} outlines the testable predictions, and Section~\ref{sec:results} presents our results.

\subsection{Experimental design}\label{sec:design}

The experiment uses a within-subject design and has two parts.
The first part of the experiment aims at identifying the participants who are regret averse.
The second part of the experiment aims at testing whether regret averse participants attempt to coordinate with their partner.

\paragraph{Part 1.}
In the first part of the experiment, participants face two decisions, {\it Decision 1} and {\it Decision 2}.
In {\it Decision 1}, participants choose between a sure amount of money (henceforth the ``sure thing'') and a risky lottery (henceforth ``the lottery'').
We selected parameters such that only an extremely risk-loving individual would prefer the lottery.
Participants who choose the lottery by-pass the body of the experiment and proceed directly to the ``Final questions'' (see below).
In {\it Decision 2}, participants who chose the sure thing must choose either to learn how much they would have earned had they chosen the lottery or not to learn, where not learning is costly.
This allows us to identify regret averse individuals, as regret averse individuals should reject the chance to learn the outcome of the unchosen option (the lottery).
{\it Decision 1} and {\it Decision 2} are further described as follows.

\bigskip

\noindent {\it Decision 1}. Participants choose between a sure amount of \textsterling3.50 and a lottery paying \textsterling10.00 with 20\% probability and \textsterling0.00 with 80\% probability. 
While the sure thing pays \textsterling3.50 for certain, the lottery has lower expected payoff of \textsterling2.00 and also brings with it substantial risk.
Our selection of parameters is designed to induce as many participants as possible to choose the sure thing over the lottery.

In practice, {\it Decision 1} proceeds as follows.
To facilitate comprehension of the lottery, we show participants an image of a jar containing five balls, four of which are blue and the remaining one is red.
We inform participants that if they choose the lottery then a ball will be drawn from the jar.
If the ball drawn is blue, they will earn \textsterling0.00; if the ball drawn is red, they will earn \textsterling10.00.
To further ensure that participants understand the relative odds of each outcome, they are required to do 100 practice draws.\footnote{Each click of the mouse performs one draw, with every ball drawn being displayed on the screen, one at a time. We opted for 100 practice draws as the empirical frequency of blue balls is statistically likely to be close to the population average of 0.8. }
Upon completion of the 100 practice draws, each participant must draw one last ball that will decide the lottery outcome, and will be hidden in a box (i.e., the participant is unable to observe the colour of this last ball).
Then each participant is asked to choose between the sure thing (\textsterling3.50) and the lottery (the unknown amount of money corresponding to the ball in the box).

\bigskip

\noindent {\it Decision 2}. Participants who chose the sure thing in {\it Decision 1} must decide whether they want to learn ``the colour of the ball in the box'' (i.e., the outcome of the foregone alternative, that in this case is how much they would have earned had they chosen the lottery).\footnote{Recall that participants who chose the lottery in {\it Decision 1} are directly sent to some final questions and then informed of their payment.}
It is not costless to avoid learning the outcome of the lottery.
A participant who decides not to learn must replicate a tedious password of 20 random characters in a text box.
Requiring that the password is typed creates a small (effort) cost, that helps us rule out the possibility that participants decide to avoid information simply because they are indifferent.
Those who pay the effort cost and choose not to learn the outcome of the lottery are classified as {\it regret averse}; those who choose to learn the outcome of the lottery are classified as {\it non regret averse}.

Between {\it Decision 1} and {\it Decision 2} participants are asked two non-incentivised questions that have no impact on their earnings in the experiment.
One question asks how they would feel if they found out that the red ball worth \textsterling10.00 was drawn,
%(possible answers: ``not pleased'' and ``indifferent'')
and the other question asks how they would feel if they found out that a blue ball worth \textsterling0.00 was drawn.
% (possible answers: ``pleased'' and ``indifferent'').
The questions are designed to help participants understand the implications of receiving information about the foregone alternative.

\paragraph{Part 2.}
In the second part of the experiment, participants take one decision, {\it Decision 3}, and then we elicit first order beliefs. At the start of the second part, participants are paired up. We ran our initial sessions with homogeneous pairs (regret averse participants were paired with other regret averse participants, and non regret averse participants with other non regret averse participants) and then further sessions with heterogeneous pairs (regret averse participants were paired with non regret averse participants).\footnote{As there were more non regret averse participants than regret averse participants, any non regret averse participant who could not be paired with a regret averse participant in the heterogenous pairs-sessions was paired with another non regret participant.} We inform participants about their partner's previous decisions.\footnote{Participants who are paired with a regret averse participant know that their partner chose the sure thing in {\it Decision 1} and chose not to learn the lottery's outcome in {\it Decision 2}.
Similarly, participants who are paired with a non regret averse participant know that their partner chose the sure amount in {\it Decision 1} and chose to learn the lottery's outcome in {\it Decision 2}.}
This information is provided to make the experimental environment closer to the theoretical model.
{\it Decision 3} is a variant of the regret game and it is used to test our model's main prediction that appears as Theorem \ref{thm:Nashmany} (case \ref{condition:Nashmany3}). The behaviour of non regret averse participants is used as a benchmark to evaluate the behaviour of regret averse participants. 
After {\it Decision 3}, all participants are asked to guess their partner's behaviour ({\it Belief}).
{\it Decision 3} and {\it Belief} are further described in the following. 

\bigskip

\noindent {\it Decision 3}. This decision is very similar to {\it Decision 2}, but has one key difference. A participant will not learn about their own lottery's outcome only if also their partner decides not to learn their own lottery's outcome. 
The same rule applies to their partner.

We emphasise that a separate lottery is drawn for each individual participant, so participants are aware that the outcome of their lottery may differ from that of their partner. This removes the potential confound of conformism.
Participants are made aware that either {\it Decision 2} or {\it Decision 3}, but not both, will be randomly selected and implemented for payment.
The reason for this is that the consequences of the two decisions may be intertwined.\footnote{For example, we wish to avoid the situation where a subject chooses not to learn the lottery's outcome in {\it Decision 2} but then chooses to learn the outcome in {\it Decision 3}, since in this case the outcome of the choice made in {\it Decision 3} overrides that made in {\it Decision 2}.} As further discussed in subsection \ref{discuss}, while {\it Decision 3} implements a variant of the regret game, the decision's incentives remain the same as in the original game. Participants play the regret game ({\it Decision 3}) only once.\footnote{The reasons why we implemented the regret game once are the following. First, we wanted to avoid repeated-game effects and particularly the effect of conforming with the partner's previous decisions. Second, given that the decision was very simple and built on {\it Decision 2}, it did not appear necessary to provide participants with learning opportunities. Third, we thought that the effect of regret aversion on behaviour may be more salient when the game is played only once.}

\bigskip

\noindent {\it Belief}.
Participants are asked to guess whether their partner chose to find out the lottery's outcome or not.
This is to determine if participants are best-responding to their beliefs.
Only 10\% of the participants are randomly selected to be paid for their guess.
The randomly selected participants earn an additional \textsterling0.50 if their guess is correct.

\paragraph{Final questions}

Before concluding each session, we ask every participant some final questions to measure risk aversion, curiosity, and demographic characteristics (age, gender, student status, and employment status).
To measure risk aversion, we endow participants with \textsterling1.00 and ask them to assign some portion of this amount to be invested in a risky venture.
If the venture is successful, participants receive 2.5 times the amount they invested; if the venture is unsuccessful, participants lose the entire amount invested.
The risky venture has the distribution of a fair coin. It is equally likely to be successful or unsuccessful.
As discussed later, curiosity is measured due to its potential interplay with the forces of regret aversion \cite{vanDijkZeelenberg:2007}.
To measure curiosity, we ask participants if they want to know the age of their now-former partner and if they want to know the outcome of their now-former partner's lottery.

\paragraph{Procedure} The experimental sessions were run online in June and July 2022 and in February 2023 using Prolific, an online participant recruitment platform.
The experiment was programmed with the software o-Tree \cite{ChenSchongerWickens}.

Every participant received a show up fee of \textsterling0.60 for participating in the experiment and a payment based on their decisions in the experiment. For those who took part in the full experiment, the average payment was \textsterling4.6 and the experiment took about 20 minutes, which corresponds to an average hourly payment of \textsterling13.8.

\subsection{Testable predictions}\label{sec:predictions}

According to our theory, in {\it Decision 3}  for a regret averse individual it is optimal to coordinate with their partner.
That is, if a regret averse individual believes their partner will choose not to learn the outcome of the foregone alternative, then they will too. If they believe that their partner will choose to learn the outcome of the foregone alternative, then they should follow suit. This is summarised by the following testable prediction.

\begin{pred}\label{pred:Reg}
Regret averse participants will attempt to coordinate on the same action.
That is, they will choose the option that they believe their partner chose.
\end{pred}

Non regret averse participants face {\it Decision 3} too, but their expected behaviour differs. In fact, for non regret averse participants, it is a dominant strategy {\it to learn} the outcome of the foregone alternative. Thus, they will choose to learn the outcome of the foregone alternative independent of their beliefs about the choice of their partner. This is summarised by the following prediction.

\begin{pred}\label{pred:NonReg}
Non regret averse participants will choose to learn the lottery's outcome independent of their beliefs about their partner's decision.
\end{pred}

\subsection{Results}\label{sec:results}

In this section, we test the predictions outlined in section \ref{sec:predictions}. 

\paragraph{Sample}~\\
As mentioned above, the identification of regret averse participants relies on asking participants whether they want to know the outcome of the foregone alternative. Thus, it is necessary that participants choose the sure thing over the lottery in {\it Decision 1}. Out of the 943 participants who started the experiment, 827 (88\%) chose the sure amount and 116 (12\%) chose the lottery. Those who chose the lottery were immediately directed to the ``Final questions'' and informed about their payment.
Those who chose the sure amount proceeded to the rest of the experiment (as described in section \ref{sec:design}). 

Of the participants who chose the sure amount in {\it Decision 1}, 757 reached {\it Decision 2}: 110 (14.5\%) chose not to learn the lottery's outcome and are classified as {\it regret averse} and 647 (85.5\%) chose to learn the lottery's outcome and are classified as {\it non regret averse}.\footnote{Being run on Prolific, this experiment was vulnerable to the potential of participants dropping out. Dropouts were mostly happening at the start of the experiment, as participants had to wait a little before being matched with another participant and they may have been unwilling to wait.} Our main analysis is based on the subsample of regret averse participants.

The participants who completed the full experiment, including the demographics questions, were 715. Of these, 61.5\% were female, 37.2\% were male and the remaining 1.3\% classified themselves as ``other (non-binary, agender and something else)''. About 71\% of the participants were employed; 14\% were students.
The average age was 40.

\paragraph{Behaviour and beliefs of regret averse participants}~\\
Table \ref{tab:Frequencies_RegAgg} reports the frequencies of {\it Decision 3} by {\it Belief} for regret averse participants.\footnote{The total number of regret averse participants in Table \ref{tab:Frequencies_RegAgg} is 107 as some participants dropped out before {\it Decision 3} or before the elicitation of beliefs.}
We find that 77\% of the regret averse participants chose not to learn the lottery's outcome in the game, and 23\% chose to learn. 
The modal action depends on participants' beliefs about their partner's choice. Almost all the participants who believed that their partner chose not to learn, also chose not to learn (56 out of 61, i.e., 92\%). That is, they attempt to coordinate with their partner, as predicted. 

\begin{table}[ht!]\centering
\def\sym#1{\ifmmode^{#1}\else\(^{#1}\)\fi}
\caption{Behaviour and beliefs of regret averse participants}\label{tab:Frequencies_RegAgg}
\setlength\tabcolsep{10pt}
\centerline{
{
\begin{tabular}{c|cc|c}
\hline
                    & Believes partner   & Believes partner   & Total   \\
                    &  does not learn  &  learns  &   \\
\hline
Does not learn             &   56  &    26 & 82\\
Learns           & 5    & 20  & 25 \\           
  \hline
Total        & 61  & 46  & 107  \\
\hline
\end{tabular}
}
}
\end{table}

About half of the participants who believed that their partner chose to learn, also chose to learn (20 out of 46, i.e., 43\%), even if learning is not what they chose in isolation. This indicates that they switched from their preferred option (not learning) to the alternative option (learning) to try and coordinate with their partner. Note that the fraction of participants who switch from not learning to learning under the belief that their partner learns (43\%) is significantly higher than the fraction of participants who switch from not learning to learning under the belief that their partner does not learn (8\%, Z-test, $p<0.0001$).

The other half of the participants who believed that their partner chose to learn, chose not to learn instead. The intuition behind this behaviour is the following. When regret averse participants believe that their partner chose not to learn, they obviously prefer not to learn, so as to avoid potential regret. When, instead, they believe that their partner chose to learn, it is not so obvious what they will want to do. Suppose that a regret averse participant believes that with a high probability their partner chose to learn and with a smaller probability their partner chose not to learn. As our experimental design elicits \emph{point} beliefs, this participant would report that they expect their partner to learn.\footnote{We elicited point beliefs rather than continuous beliefs to make the experiment easier for participants to understand.} If the cost of avoiding information is small (as it is in our experiment), it may be rational to choose not to learn. In fact, if they choose to learn they will get information for sure, which they do not like. If they choose not to learn, with some probability they will pay a small cost to avoid information but learn nevertheless, and with some probability they will manage to avoid information. 
It is reasonable to expect that, if the cost of avoiding information were individual-specific, participants' attempt to coordinate on learning would be higher. However, this calibration would not be possible in the lab. Overall, we observe that a large fraction of regret averse participants attempt to coordinate with their partner, which supports our Prediction \ref{pred:Reg}.

\begin{result} When regret averse participants believe that their partner chose not to learn, almost all of them choose not to learn. That is, they attempt to coordinate with their partner. When they believe that their partner chose to learn, nearly half of them choose to learn, even if this is not their preferred choice in isolation. 
\end{result}

\paragraph{Behaviour and beliefs of non regret averse participants}~\\
Table \ref{tab:Frequencies_NonRegAgg} reports the frequencies of {\it Decision 3} by {\it Belief} for non regret averse participants.\footnote{The total number of non regret averse participants in Table \ref{tab:Frequencies_NonRegAgg} is 609 as some participants dropped out before {\it Decision 3} or before the elicitation of beliefs.}
 Almost all the non regret averse participants (569 out of 609, i.e. over 93\%) chose to learn the lottery's outcome.  
Unlike for regret averse participants, non regret averse participants' modal action does not depend on beliefs. Nearly all the participants who expected their partner to learn, also chose to learn, (521 out of 538, i.e. 97\%).\footnote{This should not be interpreted as a coordination attempt, as for non regret averse participants choosing to learn is a dominant strategy.} Over two thirds of the participants who expected their partner not to learn, chose to learn (48 out of 71, i.e. 68\%). This indicates that learning the outcome of the foregone alternative is a dominant strategy for non regret averse participants, which supports Prediction \ref{pred:NonReg}.

\begin{table}[ht!]\centering
\def\sym#1{\ifmmode^{#1}\else\(^{#1}\)\fi}
\caption{Behaviour and beliefs of non regret averse participants}\label{tab:Frequencies_NonRegAgg}
\setlength\tabcolsep{10pt}
\centerline{
{
\begin{tabular}{c|cc|c}
\hline
                    & Believes partner   & Believes partner   & Total   \\
                    &  does not learn  &  learns  &   \\
\hline
Does not learn             &   23  &    17 & 40\\
Learns           & 48    & 521  & 569 \\           
  \hline
Total        & 71  & 538  & 609  \\
\hline
\end{tabular}
}
}
\end{table}

\begin{result} Non regret averse participants choose to learn the lottery's outcome independent of their beliefs about their partner's decision.
\end{result}

It is interesting to observe that, among the non regret averse participants who expected their partner not to learn, nearly one third (23 out of 71) also chose not to learn. This is likely to be due to other-regarding preferences. If a non regret participant expects their partner to decide not to learn, then they may decide to coordinate on information avoidance in order not to harm their partner. 

\paragraph{Behaviour and beliefs by partner type}~\\
When we disaggregate regret averse participants' behaviour and beliefs depending on whether the partner is regret averse or not, we observe that participants' beliefs change with their partner type in the right direction. Table \ref{tab:Frequencies_RegDisagg} shows regret averse participants' behaviour by beliefs, disaggregated by partner's type. In the top panel each participant's partner is also regret averse, whereas in the bottom panel the partner is non regret averse.
When regret averse participants are paired with other regret averse participants and informed that their partner also chose not to learn in the individual decision, 87\% of them expect their partner not to learn in the game. When regret averse participants are paired with non regret averse participants and informed that their partner chose to learn in the individual decision, 32\% of people expect their partner not to learn in the game. This may be driven by second order beliefs. Regret averse participants know that their (non regret averse) partner knows that they are facing someone who chose to learn in the individual decision. As a result, they may expect their (non regret averse) partner to choose not to learn in order not to hurt them.

We can also observe that beliefs affect behaviour more than partner type does. When regret averse participants believe that their partner does not learn, their behaviour is very similar regardless of their partner's type. The percentage of participants attempting coordination is 93\% when the partner is regret averse and 89\% when the partner is non regret averse.  

\bigskip

\begin{table}[ht]
  \centering
  \caption{Behaviour and beliefs by partner type for regret averse participants}\label{tab:Frequencies_RegDisagg}

  \medskip
  
\scalebox{0.9}{
  \begin{tabularx}{\linewidth}{ X X X X}
    \multicolumn{4}{c}{\textbf{Partner: regret averse}} \\
    \toprule
     & Believes partner & Believes partner & Total \\
       & does not learn & learns & \\
    \hline
    Does not learn & 39 & 2 & 41 \\
    Learns & 3 & 4 & 7 \\
    \hline
  Total        & 42  & 6  & 48  \\
    \bottomrule
  \end{tabularx}}

  \bigskip

\scalebox{0.9}{
  \begin{tabularx}{\linewidth}{ X X X X }
    \multicolumn{4}{c}{\textbf{Partner: non regret averse}} \\
    \toprule
       & Believes partner & Believes partner & Total \\
       & does not learn & learns &  \\
    \hline
Does not learn             &   17  &    24 & 41\\
Learns           & 2    & 16  & 18 \\           
  \hline
Total        & 19  & 40  & 59  \\
    \bottomrule
  \end{tabularx}}
\end{table}

In Table \ref{tab:Frequencies_NonRegDisagg} we disaggregate non regret averse participants' behaviour and beliefs by partner's type.\footnote{When we disaggregate the data for non regret averse participants into homogeneous and heterogeneous pairs, we lose a few observations because some participants' partners drop out before the end of the experiment.} In the top-panel each participant's partner is regret averse, whereas in the bottom panel the partner is non regret averse. 
When non regret averse participants are paired with other non regret averse participants and informed that their partner also chose to learn in the individual decision, 93\% of them expect their partner to learn in the game. When non regret averse participants are paired with regret averse participants and informed that their partner chose not to learn in the individual decision,  47\% of people expect their partner to learn in the game.  This may be due to the fact that these participants anticipate their (regret averse) partner's optimal decision, which is to learn. 

Also for non regret averse participants we observe that beliefs affect behaviour more than partner type does. When non regret averse participants believe that their partner does not learn, their behaviour is very similar regardless of their partner's type. When they believe that their partner chose to learn, they choose to learn with a frequency of 100\% when their partner is regret averse and with a frequency of 97\% when their partner is non regret averse.  
When they believe that their partner chose not to learn, we observe deviations from their optimal decision in isolation: they choose not to learn with a frequency of 23\% when their partner is regret averse and with a frequency of 39\% when their partner is non regret averse.

\bigskip

\begin{table}[ht]
 \centering
  \caption{Behaviour by belief and partner type for non regret averse participants}\label{tab:Frequencies_NonRegDisagg}

  \medskip
  
\scalebox{0.9}{
  \begin{tabularx}{\linewidth}{ X X X X}
    \multicolumn{4}{c}{\textbf{Partner: regret averse}} \\
    \toprule
     & Believes partner & Believes partner & Total \\
       & does not learn & learns & Total \\
    \hline
Does not learn             &   7  &    0 & 7\\
Learns           & 23    & 27  & 50 \\           
  \hline
Total        & 30  & 27  & 57  \\
    \bottomrule
  \end{tabularx}}

  \bigskip

\scalebox{0.9}{
  \begin{tabularx}{\linewidth}{ X X X X }
    \multicolumn{4}{c}{\textbf{Partner: non regret averse}} \\
    \toprule
       & Believes partner & Believes partner & Total \\
       & does not learn & learns & Total \\
    \hline
Does not learn             &   16  &    16 & 33\\
Learns           & 25    & 494  & 519 \\           
  \hline
Total        & 41  &  511 & 552  \\
    \bottomrule
  \end{tabularx}}
\end{table}

\subsection{Discussion}\label{discuss}

\paragraph{Measurement of regret aversion.} As described in section \ref{sec:design}, we classify participants as regret averse if they choose the sure thing and decide not to learn the outcome of the foregone alternative.
However, it should be noted that, besides anticipated regret aversion, there may be another psychological motive affecting people's decision to learn about unchosen options: curiosity, i.e., an intrinsic preference for non-instrumental information.

%REJOICE
%A first potential factor is rejoice. Rejoice is the psychological gain that a decision maker -- so called rejoice lover -- experiences when the unchosen option turns out to be worse than the chosen option. If anticipated rejoice is stronger than anticipated regret, a decision maker will want to know the outcome of an unchosen option.

As documented by \citet{vanDijkZeelenberg:2007}, curiosity and regret aversion are forces that pull in opposite directions.
The desire to learn what might have been may dominate the potential regret induced by learning of a bad outcome.\footnote{We believe that this force may be stronger for situations where the stakes are low.}
% this may be particularly true for low
% a decision maker's curiosity about uncertain outcomes may overcome their regret aversion and their willingness to avoid regret-inducing information.
To account for this, at the end of the study we measure participants' curiosity by asking whether they want to know the age of their now-former partner and if they want to know the outcome of their now-former partner's lottery.
Our data show that the proportion of participants classified as curious (using either question) is significantly lower among regret averse participants than among non regret averse participants.\footnote{The percentage of participants wanting to know their former partner's age was 25\% among regret averse participants and 50\% among non regret averse participants (two-sample test of proportions, $p<0.0001$). The percentage of participants wanting to know their former partner's lottery outcome was 20\% among regret averse participants and 57\% among non regret averse participants (two-sample test of proportions, $p<0.0001$).} 

This is interesting, as it suggests that the 14.5\% of participants who decided to avoid potentially regret-inducing information in our study are not likely to be {\it the only} regret averse participants in our sample.
They are those whose regret aversion dominates curiosity.
Among the participants who decided to receive potentially regret-inducing information in our study there will have been participants who experienced anticipated regret aversion, but whose anticipated regret aversion was not sufficiently strong to overcome the desire to find out.
%\footnote{In a previous experiment \cite{CerroneFeriNearyWP} we used a very similar design where the cost of avoiding information was monetary rather than effort-based. We found an almost identical percentage of regret averse participants (11.7\%). This suggests that our results are robust.} 
The design of a measure of regret aversion that can fully separate regret from curiosity is left for future research.

% SHOULD WE MENTION THE METHOD USED IN EXPERIMENT 1?
% In a previous experiment \citet{CerroneFeriNearyWP}, we elicit participants' preferences over a sure amount of money and a lottery under two feedback conditions -- one where participants learn the lottery's outcome even if they do not choose it and one where they do not learn the lottery's outcome unless they choose it.\footnote{The idea that the expectation of feedback can affect the behaviour of regret averse individuals has been first documented in the psychology literature. \citet{Zeelenberg:1996} find that, when faced with the choice between two equally attractive gambles, most participants chose the gamble without ex-post feedback.} Participants who find the sure amount more appealing when the outcome of the lottery is not revealed are classified as regret averse -- by choosing the sure amount they can avoid information about the unchosen option and thus potential regret.\footnote{In a similar fashion, \citet{ImasLameWilson} compare participants' valuations for identical lotteries under two different feedback scenarios. However, they use a between-subject design and it is not possible to back out the number of regret averse individuals from their data.}

\paragraph{Implementation of the regret game.}
As described in section \ref{sec:design}, we implement the regret game by asking participants whether they want to learn the outcome of the foregone alternative, where they can avoid learning only if their partner chooses the same ({\it Decision 3}). The straight way to implement the regret game (as descibed in our theoretical model) would be to ask participants to choose between a sure amount of money and a lottery, where if they do not choose the lottery, they learn its outcome if their partner chooses the lottery.
However, such an implementation, used in a previous version of this paper, is vulnerable to two potential confounds. First, when participants do not know ex ante what the best choice is (given that the lottery is risky), they might choose what they believe their partner chose because they do not know what is best and thus prefer to try and conform with others \cite{CharnessRigottiRustichini:2017:GEB,CharnessNaefSontuoso:2019:JET}. Second, when participants' decisions determine their earnings, they might choose what they believe their partner chose because they do not want to earn less than their partner \cite{FehrSchmidt:1999:QJE}. In the current experiment these two potential confounds are eliminated by design. First, there is no uncertainty in {\it Decision 3} and thus participants can immediately identify the best decision to take given their preferences -- they have no reason to desire to conform. Second, {\it Decision 3} does not affect participants' earnings and thus participants have no reason to imitate their partner to avoid earning less.

Note that in the implemented regret game, when the cost of avoiding information is lower than the expected regret, we have a coordination game, where the Pareto efficient equilibrium (both players choose not to learn) is equivalent to the equilibrium where all players choose the risk-free lottery in Theorem \ref{thm:Nashmany} .

\section{Conclusion}\label{sec:conclusion}

%This paper began with the following simple observation:\ in many situations, ranging from technology adoption to ordering food in a restaurant, learning the outcome of unchosen alternatives is not guaranteed.

%Given that in formal models of regret averse decision making an individual can only regret a foregone alternative if its outcome is learned, how should these models be amended?
%We show how to extend existing models of regret averse decision making to incorporate this observation by enriching the notion of ``ex-post information''.
%We then show that a ``more informative'' ex-post information environment is never preferred for a regret-averse decision maker.
%At least for those who suffer from regret, the old adage is true:\ ignorance is bliss.

This paper began with the following simple observation:\ in many situations, ranging from technology adoption to ordering food in a restaurant, learning the outcome of unchosen alternatives is not guaranteed.
While being blissfully ignorant is desirable for a regret averse decision maker, the problem, as so often in life, is others.
Others can make a choice that you did not make and that then turned out to have a better realisation than the choice you made.
%We model this as the possibility that the ex-post information environment that a regret averse decision maker will face depends not only on their own choice but also on the choices of others.
We model this by allowing the ex-post information environment faced by a collection of regret averse decision makers to depend not only on their own choice but also on the choices of others.
Thus, what appeared to be a series of individual decision problems - like for example ordering food when dining with friends - becomes a rich multi-player behavioural game.
The reason is that the decisions of others - what one's friends order - can be informative about foregone alternatives, and for a regret averse individual that can be impactful.
We term this environment the {\it regret game}, and classify the conditions on preferences for which it is a coordination game with multiple equilibria.

We test our model's predictions through an experiment. The experiment's goal is to test whether regret averse participants attempt to coordinate with their partner as our model predicts.
In the first part of the experiment, we identify regret averse participants by asking them whether they want to learn about a foregone alternative or not, where avoiding information involves a small effort cost. Those who choose to avoid information and pay the associated cost are classified as {\it regret averse}. In the second part of the experiment, we pair up participants.
Again, individuals decide whether they want to learn about the foregone alternative or not, but avoiding information now also depends on their partner's decision:\ they will not learn about the foregone alternative only if also their partner decides not to learn about it. 
We find strong evidence that, consistent with our predictions, regret averse participants attempt to coordinate with their partner. 
%Regret averse participants' modal action depends on their beliefs. The frequency with which they choose the action that they believe their partner chose is significantly higher than the frequency with which they choose the alternative action. Finally, regret aversion significantly increases participants' probability of choosing not to learn about the foregone alternative under belief that their partner also chooses not to learn. 

%\newpage 

The general topic of how individual psychological motives can lead to socially interdependent decisions is, in our opinion, rather underexplored.
While we have focused on regret averse decision makers, our contribution speaks to the broader theme of how the behaviour of others can shape one's informational environment.
%how the ex-post information available to individuals may depend on the decisions of others.
Given that ex-post information matters for individuals with certain behavioural biases, there are strategic interdependencies at play.
We hope that our paper will stimulate further theoretical and experimental work to understand how individuals with behavioural preferences are connected through information.

\newpage

\bibliographystyle{elsarticle-harv}
\bibliography{regret}

\begin{thebibliography}{33}
\expandafter\ifx\csname natexlab\endcsname\relax\def\natexlab#1{#1}\fi
\providecommand{\url}[1]{\texttt{#1}}
\providecommand{\href}[2]{#2}
\providecommand{\path}[1]{#1}
\providecommand{\DOIprefix}{doi:}
\providecommand{\ArXivprefix}{arXiv:}
\providecommand{\URLprefix}{URL: }
\providecommand{\Pubmedprefix}{pmid:}
\providecommand{\doi}[1]{\href{http://dx.doi.org/#1}{\path{#1}}}
\providecommand{\Pubmed}[1]{\href{pmid:#1}{\path{#1}}}
\providecommand{\bibinfo}[2]{#2}
\ifx\xfnm\relax \def\xfnm[#1]{\unskip,\space#1}\fi
%Type = Article
\bibitem[{Ariely and Levav(2000)}]{ArielyLevav:2000:JCR}
\bibinfo{author}{Ariely, D.}, \bibinfo{author}{Levav, J.},
  \bibinfo{year}{2000}.
\newblock \bibinfo{title}{Sequential choice in group settings: Taking the road
  less traveled and less enjoyed}.
\newblock \bibinfo{journal}{Journal of Consumer Research} \bibinfo{volume}{27},
  \bibinfo{pages}{279--290}.
%Type = Article
\bibitem[{Bandiera and Rasul(2006)}]{BandieraRasul:2006:EJ}
\bibinfo{author}{Bandiera, O.}, \bibinfo{author}{Rasul, I.},
  \bibinfo{year}{2006}.
\newblock \bibinfo{title}{Social networks and technology adoption in northern
  mozambique}.
\newblock \bibinfo{journal}{The Economic Journal} \bibinfo{volume}{116},
  \bibinfo{pages}{869--902}.
%Type = Article
\bibitem[{Banerjee et~al.(2013)Banerjee, Chandrasekhar, Duflo and
  Jackson}]{BanerjeeChandrasekhar:2013:S}
\bibinfo{author}{Banerjee, A.}, \bibinfo{author}{Chandrasekhar, A.G.},
  \bibinfo{author}{Duflo, E.}, \bibinfo{author}{Jackson, M.O.},
  \bibinfo{year}{2013}.
\newblock \bibinfo{title}{The diffusion of microfinance}.
\newblock \bibinfo{journal}{Science} \bibinfo{volume}{341}.
%Type = Article
\bibitem[{Banerjee(1992)}]{Banerjee:1992:QJE}
\bibinfo{author}{Banerjee, A.V.}, \bibinfo{year}{1992}.
\newblock \bibinfo{title}{A simple model of herd behavior}.
\newblock \bibinfo{journal}{The Quarterly Journal of Economics}
  \bibinfo{volume}{107}, \bibinfo{pages}{797--817}.
%Type = Unpublished
\bibitem[{Battigalli et~al.(2022)Battigalli, Dufwenberg and
  Li}]{BattigalliDufwenbergLi}
\bibinfo{author}{Battigalli, P.}, \bibinfo{author}{Dufwenberg, M.},
  \bibinfo{author}{Li, S.}, \bibinfo{year}{2022}.
\newblock \bibinfo{title}{Regret in games}.
%Type = Article
\bibitem[{Bell(1982)}]{Bell:1982}
\bibinfo{author}{Bell, D.E.}, \bibinfo{year}{1982}.
\newblock \bibinfo{title}{Regret in decision making under uncertainty}.
\newblock \bibinfo{journal}{Operations Research} \bibinfo{volume}{30},
  \bibinfo{pages}{961--981}.
%Type = Article
\bibitem[{B{\'e}nabou(2013)}]{Benabou:2013:RES}
\bibinfo{author}{B{\'e}nabou, R.}, \bibinfo{year}{2013}.
\newblock \bibinfo{title}{Groupthink: Collective delusions in organizations and
  markets}.
\newblock \bibinfo{journal}{The Review of Economic Studies}
  \bibinfo{volume}{80}, \bibinfo{pages}{429--462}.
%Type = Article
\bibitem[{Bikhchandani et~al.(1992)Bikhchandani, Hirshleifer and
  Welch}]{BikhchandaniHirshleifer:1992:JPE}
\bibinfo{author}{Bikhchandani, S.}, \bibinfo{author}{Hirshleifer, D.},
  \bibinfo{author}{Welch, I.}, \bibinfo{year}{1992}.
\newblock \bibinfo{title}{A theory of fads, fashion, custom, and cultural
  change as informational cascades}.
\newblock \bibinfo{journal}{Journal of Political Economy}
  \bibinfo{volume}{100}, \bibinfo{pages}{pp. 992--1026}.
%Type = Article
\bibitem[{Carlsson and Damme(1993)}]{CarlssonDamme:1993:E}
\bibinfo{author}{Carlsson, H.}, \bibinfo{author}{Damme, E.v.},
  \bibinfo{year}{1993}.
\newblock \bibinfo{title}{Global games and equilibrium selection}.
\newblock \bibinfo{journal}{Econometrica} \bibinfo{volume}{61},
  \bibinfo{pages}{989--1018}.
%Type = Article
\bibitem[{Charness et~al.(2019)Charness, Naef and
  Sontuoso}]{CharnessNaefSontuoso:2019:JET}
\bibinfo{author}{Charness, G.}, \bibinfo{author}{Naef, M.},
  \bibinfo{author}{Sontuoso, A.}, \bibinfo{year}{2019}.
\newblock \bibinfo{title}{Opportunistic conformism}.
\newblock \bibinfo{journal}{Journal of Economic Theory} \bibinfo{volume}{180},
  \bibinfo{pages}{100--134}.
%Type = Article
\bibitem[{Charness et~al.(2017)Charness, Rigotti and
  Rustichini}]{CharnessRigottiRustichini:2017:GEB}
\bibinfo{author}{Charness, G.}, \bibinfo{author}{Rigotti, L.},
  \bibinfo{author}{Rustichini, A.}, \bibinfo{year}{2017}.
\newblock \bibinfo{title}{Social surplus determines cooperation rates in the
  one-shot prisoner's dilemma}.
\newblock \bibinfo{journal}{Games and Economic Behavior} \bibinfo{volume}{100},
  \bibinfo{pages}{113--124}.
%Type = Article
\bibitem[{Chen et~al.(2016)Chen, Schonger and Wickens}]{ChenSchongerWickens}
\bibinfo{author}{Chen, D.L.}, \bibinfo{author}{Schonger, M.},
  \bibinfo{author}{Wickens, C.}, \bibinfo{year}{2016}.
\newblock \bibinfo{title}{otree -- an open-source platform for laboratory,
  online, and field experiments}.
\newblock \bibinfo{journal}{Journal of Behavioral and Experimental Finance}
  \bibinfo{volume}{9}, \bibinfo{pages}{88--97}.
%Type = Article
\bibitem[{Conley and Udry(2010)}]{ConleyUdry:2010:AER}
\bibinfo{author}{Conley, T.G.}, \bibinfo{author}{Udry, C.R.},
  \bibinfo{year}{2010}.
\newblock \bibinfo{title}{Learning about a new technology: Pineapple in ghana}.
\newblock \bibinfo{journal}{American Economic Review} \bibinfo{volume}{100},
  \bibinfo{pages}{35--69}.
%Type = Article
\bibitem[{Cooper and Rege(2011)}]{CooperRege:2011:GEB}
\bibinfo{author}{Cooper, D.J.}, \bibinfo{author}{Rege, M.},
  \bibinfo{year}{2011}.
\newblock \bibinfo{title}{Misery loves company: Social regret and social
  interaction effects in choices under risk and uncertainty}.
\newblock \bibinfo{journal}{Games and Economic Behavior} \bibinfo{volume}{73},
  \bibinfo{pages}{91--110}.
%Type = Article
\bibitem[{Crawford(1991)}]{Crawford:1991:GEB}
\bibinfo{author}{Crawford, V.P.}, \bibinfo{year}{1991}.
\newblock \bibinfo{title}{An ``evolutionary'' interpretation of van huyck,
  battalio, and beil's experimental results on coordination}.
\newblock \bibinfo{journal}{Games and Economic Behavior} \bibinfo{volume}{3},
  \bibinfo{pages}{25--59}.
%Type = Article
\bibitem[{DeGroot(1974)}]{DeGroot:1974:JASA}
\bibinfo{author}{DeGroot, M.H.}, \bibinfo{year}{1974}.
\newblock \bibinfo{title}{Reaching a consensus}.
\newblock \bibinfo{journal}{Journal of the American Statistical Association}
  \bibinfo{volume}{69}, \bibinfo{pages}{118--121}.
%Type = Article
\bibitem[{van Dijk and Zeelenberg(2007)}]{vanDijkZeelenberg:2007}
\bibinfo{author}{van Dijk, E.}, \bibinfo{author}{Zeelenberg, M.},
  \bibinfo{year}{2007}.
\newblock \bibinfo{title}{When curiosity killed regret: Avoiding or seeking the
  unknown in decision making under uncertainty}.
\newblock \bibinfo{journal}{Journal of Experimental Social Psychology}
  \bibinfo{volume}{43}, \bibinfo{pages}{656--662}.
%Type = Article
\bibitem[{Fehr and Schmidt(1999)}]{FehrSchmidt:1999:QJE}
\bibinfo{author}{Fehr, E.}, \bibinfo{author}{Schmidt, K.M.},
  \bibinfo{year}{1999}.
\newblock \bibinfo{title}{A theory of fairness, competition, and cooperation}.
\newblock \bibinfo{journal}{The Quarterly Journal of Economics}
  \bibinfo{volume}{114}, \bibinfo{pages}{817--868}.
%Type = Article
\bibitem[{Filiz-Ozbay and Ozbay(2007)}]{Filiz-OzbayOzbay:2007:AER}
\bibinfo{author}{Filiz-Ozbay, E.}, \bibinfo{author}{Ozbay, E.Y.},
  \bibinfo{year}{2007}.
\newblock \bibinfo{title}{Auctions with anticipated regret: Theory and
  experiment}.
\newblock \bibinfo{journal}{The American Economic Review} \bibinfo{volume}{97},
  \bibinfo{pages}{1407 -- 1418}.
%Type = Article
\bibitem[{Foster and Young(1990)}]{FosterYoung:1990:TPB}
\bibinfo{author}{Foster, D.}, \bibinfo{author}{Young, P.},
  \bibinfo{year}{1990}.
\newblock \bibinfo{title}{Stochastic evolutionary game dynamics}.
\newblock \bibinfo{journal}{Theoretical Population Biology}
  \bibinfo{volume}{38}, \bibinfo{pages}{219--232}.
%Type = Article
\bibitem[{Geanakoplos et~al.(1989)Geanakoplos, Pearce and
  Stacchetti}]{GeanakoplosPearce:1989:JET}
\bibinfo{author}{Geanakoplos, J.}, \bibinfo{author}{Pearce, D.},
  \bibinfo{author}{Stacchetti, E.}, \bibinfo{year}{1989}.
\newblock \bibinfo{title}{Psychological games and sequential rationality}.
\newblock \bibinfo{journal}{Games and Economic Behavior} \bibinfo{volume}{1},
  \bibinfo{pages}{60--79}.
%Type = Book
\bibitem[{Harsanyi and Selten(1988)}]{HarsanyiSelten:1988:}
\bibinfo{author}{Harsanyi, J.C.}, \bibinfo{author}{Selten, R.},
  \bibinfo{year}{1988}.
\newblock \bibinfo{title}{A General Theory of Equilibrium Selection in Games}.
\newblock \bibinfo{publisher}{MIT Press}.
%Type = Article
\bibitem[{Humphrey(2004)}]{Humphrey:2004:JEP}
\bibinfo{author}{Humphrey, S.J.}, \bibinfo{year}{2004}.
\newblock \bibinfo{title}{Feedback-conditional regret theory and testing
  regret-aversion in risky choice}.
\newblock \bibinfo{journal}{Journal of Economic Psychology}
  \bibinfo{volume}{25}, \bibinfo{pages}{839--857}.
%Type = Article
\bibitem[{Kahneman and Riepe(1998)}]{KahnemanRiepe:1998:JPM}
\bibinfo{author}{Kahneman, D.}, \bibinfo{author}{Riepe, M.W.},
  \bibinfo{year}{1998}.
\newblock \bibinfo{title}{Aspects of investor psychology}.
\newblock \bibinfo{journal}{The Journal of Portfolio Management}
  \bibinfo{volume}{24}, \bibinfo{pages}{52--65}.
%Type = Article
\bibitem[{Kandori et~al.(1993)Kandori, Mailath and Rob}]{KandoriMailath:1993:E}
\bibinfo{author}{Kandori, M.}, \bibinfo{author}{Mailath, G.J.},
  \bibinfo{author}{Rob, R.}, \bibinfo{year}{1993}.
\newblock \bibinfo{title}{Learning, mutation, and long run equilibria in
  games}.
\newblock \bibinfo{journal}{Econometrica} \bibinfo{volume}{61},
  \bibinfo{pages}{29--56}.
%Type = Article
\bibitem[{Kreps and Porteus(1978)}]{KrepsPorteus:1978:E}
\bibinfo{author}{Kreps, D.M.}, \bibinfo{author}{Porteus, E.L.},
  \bibinfo{year}{1978}.
\newblock \bibinfo{title}{Temporal resolution of uncertainty and dynamic choice
  theory}.
\newblock \bibinfo{journal}{Econometrica} \bibinfo{volume}{46},
  \bibinfo{pages}{185--200}.
%Type = Article
\bibitem[{Loomes and Sugden(1982)}]{LoomesSugden:1982:EJ}
\bibinfo{author}{Loomes, G.}, \bibinfo{author}{Sugden, R.},
  \bibinfo{year}{1982}.
\newblock \bibinfo{title}{Regret theory: An alternative theory of rational
  choice under uncertainty}.
\newblock \bibinfo{journal}{Economic Journal} \bibinfo{volume}{92},
  \bibinfo{pages}{805--824}.
%Type = Inproceedings
\bibitem[{Morris and Shin(2003)}]{MorrisShin:2003:}
\bibinfo{author}{Morris, S.}, \bibinfo{author}{Shin, H.S.},
  \bibinfo{year}{2003}.
\newblock \bibinfo{title}{Global games: Theory and applications}, in:
  \bibinfo{booktitle}{in ``Advances in Economics and Econometrics, the Eighth
  World Congress'', Dewatripont, Hansen and Turnovsky, Eds}.
%Type = Article
\bibitem[{Rothschild and Stiglitz(1976)}]{RothschildStiglitz:1976:QJE}
\bibinfo{author}{Rothschild, M.}, \bibinfo{author}{Stiglitz, J.E.},
  \bibinfo{year}{1976}.
\newblock \bibinfo{title}{Equilibrium in competitive insurance markets: An
  essay on the economics of imperfect information}.
\newblock \bibinfo{journal}{The Quarterly Journal of Economics}
  \bibinfo{volume}{90}, \bibinfo{pages}{630--49}.
%Type = Article
\bibitem[{Ryan and Gross(1943)}]{RyanGross:1943:RS}
\bibinfo{author}{Ryan, B.}, \bibinfo{author}{Gross, N.}, \bibinfo{year}{1943}.
\newblock \bibinfo{title}{{The diffusion of hybrid seed corn in two Iowa
  communities}}.
\newblock \bibinfo{journal}{Rural Sociology} \bibinfo{volume}{8},
  \bibinfo{pages}{15--24}.
%Type = Article
\bibitem[{Spence(1973)}]{Spence:1973:QJE}
\bibinfo{author}{Spence, M.}, \bibinfo{year}{1973}.
\newblock \bibinfo{title}{Job market signaling}.
\newblock \bibinfo{journal}{The Quarterly Journal of Economics}
  \bibinfo{volume}{87}, \bibinfo{pages}{355--374}.
%Type = Article
\bibitem[{Van~Huyck et~al.(1990)Van~Huyck, Battalio and
  Beil}]{Van-HuyckBattalio:1990:AER}
\bibinfo{author}{Van~Huyck, J.B.}, \bibinfo{author}{Battalio, R.C.},
  \bibinfo{author}{Beil, R.O.}, \bibinfo{year}{1990}.
\newblock \bibinfo{title}{Tacit coordination games, strategic uncertainty, and
  coordination failure}.
\newblock \bibinfo{journal}{American Economic Review} \bibinfo{volume}{80},
  \bibinfo{pages}{234--48}.
%Type = Article
\bibitem[{Young(1993)}]{Young:1993:E}
\bibinfo{author}{Young, H.P.}, \bibinfo{year}{1993}.
\newblock \bibinfo{title}{The evolution of conventions}.
\newblock \bibinfo{journal}{Econometrica} \bibinfo{volume}{61},
  \bibinfo{pages}{57--84}.

\end{thebibliography}
%\bibliography{regret.bib}

\newpage

\appendix

\footnotesize{

\section*{Appendix}\label{APP}

\addcontentsline{toc}{section}{Appendix}
\renewcommand{\thesubsection}{\Alph{subsection}}

\subsection{Proof of Theorem \ref{thm:Nashmany}}\label{App:proofs}

\setcounter{equation}{0}\renewcommand{\theequation}{A\arabic{equation}}

Parts \ref{condition:Nashmany1} and \ref{condition:Nashmany2} are immediate as all players have a dominant strategy over each range of parameters.
Consider part \ref{condition:Nashmany3}. It is easy to see that both symmetric profiles are Nash Equilibria over this range.
To see that the symmetric outcomes are the {only} pure strategy Nash Equilibria, suppose to the contrary. That is, suppose there is a pure strategy Nash equilibrium, $\hat{\abold}$, in which some individual, say $p$, chooses $\ell_{S}$ and another individual, say $q$, chooses the risky lottery, $\ell_{r}$. Since $\phi$ is defined as a strictly increasing function from $\set{0, 1,  {2}, \dots, N-1}$ to $[0,1]$, we have that $\phi_p(\hat{\abold}) = \phi(\#\set{r \neq p : \hat{a}_{r} = \ell }) > \phi(\#\set{r \neq q : \hat{a}_{r} = \ell}) = \phi_q(\hat{\abold})$. But this contradicts the fact that $\ell_{S}$ is optimal for individual $p$ and that $\ell$ is optimal for individual $q$, and so profile $\hat{\abold}$ cannot be a pure strategy Nash equilibrium.

\subsection{Experimental instructions}\label{instructions}
{\it The text in italics was not shown to participants -- it is added here to better describe the instructions.}

\bigskip

[Consent form]

\bigskip

\textbf{Welcome}

\smallskip

Thank you for being a participant in this study. 

\bigskip

Please give this study your full attention.
The study will involve multiple screens with each screen having a countdown timer.
If time runs out, you will be excluded from the study and receive no further payment.
This is not to put you under time pressure, but to ensure that the study is run smoothly.

\bigskip

{\it New screen}

\smallskip

There is a jar including 5 balls:\ 4 blue balls and 1 red ball.
One ball will be drawn from the jar.
If a blue ball appears, then you get \textsterling0.
If the red ball appears, then you get \textsterling10.
This means that when you draw a ball from this jar there is a 20\% chance of earning \textsterling10 and an 80\% chance of earning \textsterling0. That is, drawing a ball from this jar pays \textsterling2 on average.

\bigskip

\begin{figure}[h!]
\centering 
\includegraphics[scale=0.22]{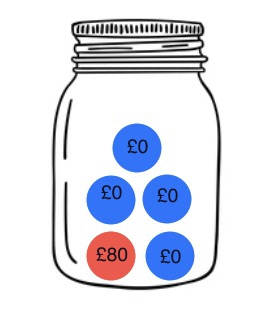}
\caption*{}
\label{fig:urn}\end{figure}

\bigskip

To help you understand these odds, you will do 100 practice draws by clicking on the button ``Draw a ball''. For each draw the odds remain the same. 

\bigskip

\mybox{Draw a ball} \hspace{0.6cm} {\it Each time this button is clicked a ball is displayed here (one at a time).} 

\bigskip

{\it After the 100th ball has been drawn, the following text appears.}

\bigskip

The 100 practice draws are over. Now draw one more ball and we will put it into a box.

\bigskip

\mybox{Draw a ball} \hspace{0.6cm} {\it When this button is clicked, a yellow box with a question mark appears, as well as the following text.}

\bigskip

The last ball drown is in the yellow box. This means that the yellow box contains a blue ball (worth \textsterling0) with 80\% probability and a red ball (worth \textsterling10) with 20\% probability.

\bigskip

For your participation so far, you have earned \textsterling3.50. Do you want to keep your \textsterling3.50 or do you want to exchange your \textsterling3.50 for the ball hidden in the box?
\begin{itemize}[label=$\circ$]
\item keep the \textsterling3.50
\vspace{-1mm}
\item exchange the \textsterling3.50 for the ball hidden in the box
\end{itemize}

\bigskip

{\it The participants who choose the ball in the box directly proceed to the screen with demographic and other final questions. For the participants who choose the \textsterling3.50, the experiment continues as follows.}

\bigskip

{\it New screen}

\smallskip

If you open the box, you will find out whether you would have earned \textsterling0 or \textsterling10, if you had exchanged your \textsterling3.50 for the ball in the box. 

\bigskip

If you found out that the box contained a red ball (worth \textsterling10), how would you feel?
\begin{itemize}[label=$\circ$]
\item Not pleased
\vspace{-1mm}
\item Indifferent
\end{itemize}

\bigskip

If you found out that the box contained a blue ball (worth \textsterling0), how would you feel?
\begin{itemize}[label=$\circ$]
\item Pleased
\vspace{-1mm}
\item Indifferent
\end{itemize}

\bigskip

Do you want to know what the box contains? 

\bigskip

\textbf{If you decide to know} how much you would have earned, in the next screen you will find out whether the box contained a blue ball (worth \textsterling0) or a red ball (worth \textsterling10).

\bigskip

\textbf{If you decide NOT to know} how much you would have earned, you have to type a password to keep the box closed.

\bigskip

What do you prefer? Please make your choice.
 \begin{itemize}[label=$\circ$]
\item I want to know how much I would have earned.\\  Click on ``Next'' to find out whether the box contained a blue ball (worth \textsterling0) or a red ball (worth \textsterling10). \mybox{Next}
\item I don't want to know how much I would have earned.\\ Type the password into the empty space to keep the box closed and then click on ``Submit''.
\vspace{-1mm}
\item [] swZHNnFzvkHaQg2zhpoE \hspace{5mm} $\underline{\hspace{3cm}}$ \hspace{5mm} \mybox{Submit}

\end{itemize}

\bigskip

Either this decision or the decision on the next page will be randomly selected and implemented.

\bigskip

{\it New screen}

\smallskip

In this part of the study, you have been randomly paired with another participant. Like you, your partner chose to keep the \textsterling3.50 and then chose not to know\footnote{For participants paired with a non regret averse partner, this sentence says ``your partner (...) chose to know''.} the colour of the ball in their box. Your partner will never know your identity and you will never know the identity of your partner.

\bigskip

As in the previous decision, you have to decide whether you want to know the colour of the ball in your box. \textbf{But there is one important difference now. If you decide not to know, your box will remain closed only if your partner decides not to know as well. The same applies to your partner. That is, if either of you chooses to know, then both of you will find out the colour of the ball in their box}.

\bigskip

Note: the ball is drawn individually for each participant. This means that the ball in your box and the ball in your partner's box may be of a different colour.

\bigskip

Please choose one of the following answers. 
\begin{itemize}[label=$\circ$]
\item I want to know how much I would have earned.\\  Click on ``Next'' to find out whether the box contained a blue ball (worth \textsterling0) or a red ball (worth \textsterling10). \mybox{Next}
\item I don't want to know how much I would have earned.\\ Type the password into the empty space to keep your box closed and then click on ``Submit''.\\
If also your partner decides not to know and types the password, each of your boxes will remain closed. If your partner decides to know, you will both find out what is in your own box regardless.
\vspace{-1mm}
\item [] YqSKxYJBSqOaD59k7pBb \hspace{5mm} $\underline{\hspace{3cm}}$ \hspace{5mm} \mybox{Submit}
\end{itemize}

\bigskip

Either this decision or the decision on the next page will be randomly selected and implemented.

\bigskip

{\it New screen}

\smallskip

Which option do you think your partner chose? 
 \begin{itemize}[label=$\circ$]
\item My partner chose to find out the amount in their box
\vspace{-1mm}
\item My partner chose not to find out the amount in their box
\end{itemize}

\bigskip

Only 10\% of the participants in this study will be randomly selected to be paid for this guess. The randomly selected participants will earn additional \textsterling0.50 if their guess is correct. 

\bigskip

{\it New screen}

\smallskip

You are now given \textsterling1 and must choose the portion of this amount (between \textsterling0 and \textsterling1, inclusive) that you wish to invest. The amount not invested is yours to keep.

\bigskip

The investment is risky. It will be successful with 50\% and unsuccessful with 50\% probability.\\ 
If the investment is successful, you receive 2.5 times the amount you chose to invest.\\If the investment is unsuccessful, you lose the amount invested.

\bigskip

How much would you like to invest? $\underline{\hspace{1cm}}$

\bigskip

Your age: $\underline{\hspace{1cm}}$

\bigskip

Your gender: [female, male, other (non-binary, agender, something else)]

\bigskip

Are you a student? [Yes, No]

\bigskip

Are you employed? [Yes, No]

\bigskip

Would you like to know the age of your partner? [Yes, No]

\bigskip

Would you like to know the colour of the ball in your partner's box? [Yes, No]

}

%\setlength\parindent{0pt}

%\marginpar{\tiny{\CC{Bla}}}

\end{document}